\documentclass[12pt,a4paper]{article}

\newcommand{\al}{\alpha}
\newcommand{\x}{\tilde{x}}
\newcommand{\bz}{\bar{z}}
\newcommand{\pa}{\partial}
\newcommand{\ga}{\gamma}
\newcommand{\vx}{\mbox{\boldmath$x$}}

\begin{document}

\begin{flushright}
{}
\end{flushright}
\vspace{1.8cm}

\begin{center}
 \textbf{\Large Asymptotic AdS String Solutions\\
for Null Polygonal Wilson Loops in $R^{1,2}$ }
\end{center}
\vspace{1.6cm}
\begin{center}
 Shijong Ryang
\end{center}

\begin{center}
\textit{Department of Physics \\ Kyoto Prefectural University of Medicine
\\ Taishogun, Kyoto 603-8334 Japan}  \par
\texttt{ryang@koto.kpu-m.ac.jp}
\end{center}
\vspace{2.8cm}
\begin{abstract}
For the asymptotic string solution in $AdS_3$ which is represented by
the $AdS_3$ Poincare coordinates and yields the planar multi-gluon 
scattering amplitude at strong coupling in arXiv:0904.0663, we 
express it by the $AdS_4$ Poincare coordinates and demonstrate that
the hexagonal and octagonal Wilson loops surrounding the string
surfaces take closed contours consisting of null vectors in $R^{1,2}$ 
owing to the relations of Stokes matrices. For the tetragonal Wilson
loop we construct a string solution characterized by two parameters
by solving the auxiliary linear problems and demanding a reality 
condition, and analyze the asymptotic behavior of the solution 
in $R^{1,2}$. The freedoms of two parameters are related with some 
conformal SO(2,4) transformations.
\end{abstract} 
\vspace{3cm}
\begin{flushleft}
October, 2009
\end{flushleft}

\newpage
\section{Introduction}

The AdS/CFT correspondence has more and more revealed the deep
relations between the $\mathcal{N}=4$ super Yang-Mills (SYM) theory
and the string theory in $AdS_5\times S^5$, where classical string 
solutions play an important role \cite{GKP,AT,JP}. 
Alday and Maldacena \cite{AM} have evaluated the
planar four-gluon scattering amplitude at strong coupling in the 
$\mathcal{N}=4$ SYM theory by computing the four-cusp Wilson loop composed
of four lightlike gluon momenta from a certain open string solution in
$AdS$ space whose worldsheet surface is related by a conformal 
SO(2,4) transformation to the one-cusp Wilson loop surface found in 
\cite{MK}. The dimensionally regularized four-gluon amplitude
at strong coupling agrees with the BDS conjectured form regarding the 
all-loop iterative structure and the IR divergence of the perturbative 
gluon amplitude \cite{BDS}.

Inspired by this investigation there have been a lot of works about
the string theory computations of the gluon amplitudes
and the constructions of open string solutions in $AdS$ with null
boundaries \cite{AFK,ADI,LAM,JKS,KRT,SR,AR}.

On the other hand based on the perturbative approach in the SYM theory
various studies have been made about duality between the planar gluon
amplitudes \cite{CSV} and the null Wilson loops where both have the same
(dual) conformal symmetry, and the anomalous conformal Ward identity which
constrains the Wilson loops \cite{DHK}.

Using the Pohlmeyer reduction \cite{KP,VS} several string 
solutions in $AdS$ have been constructed from the soliton solutions in the
generalized Sinh-Gordon model\cite{JKS,JJ}.
Within the Pohlmeyer reduction the string configurations with
the timelike and spacelike minimal surfaces in  $AdS$ \cite{DJW}
and the closed strings in SL(2,R) \cite{GJ} have been studied.

Recently the multi-gluon scattering amplitudes at strong coupling
have been investigated \cite{FAM}, where the asymptotic form
of string configuration in $AdS_3$ is constructed in the complex
worldsheet $z$ plane by using the Pohlmeyer reduction and solving
approximately the auxiliary left and right linear problems
in large $z$ involving a single field $\al$ which obeys a generalized
Sinh-Gordon equation. Through the Stokes phenomenon as 
$z \rightarrow \infty$ which is expressed by the Stokes matrices,
the various cusps for the Wilson loop appear in the various angular
sectors of the  $z$ plane. The problem to compute the minimal area for
the octagonal Wilson loop is reduced to the study of SU(2) Hitchin 
equations \cite{GMN}. The octagonal gluon amplitude has been 
computed by using the asymptotic form of the solution in large $z$
and the remainder function has been constructed to be expressed
in terms of the spacetime cross ratios.

Starting with the genus one finite-gap form for the string solution
in $AdS$, a classification of the allowed solutions has been performed
by solving the reality and Virasoro conditions \cite{SS}, where
there is a construction of a class of solutions with six null boundaries,
among which two pairs are collinear.

The asymptotic string solution for the multi-gluon amplitude in
ref. \cite{FAM} is represented by the $AdS_3$ Poincare coordinates
$(r,x^+,x^-)$, where a polygonal Wison loop is going around the
cylinder that is identified with the two dimensional space $R^{1,1}$,
namely, the $(x^+, x^-)$ space. We will regard this string 
configuration as living in $AdS_3$ subspace of $AdS_4$ and 
express it in terms of the $AdS_4$ Poincare coordinates 
$(\tilde{r},\x_0,\x_1,\x_2)$ and see how the hexagonal 
and octagonal Wilson loops at the $AdS_4$ boundary
are constructed by connecting a sequence of null vectors in a closed
form in the three dimensional space $R^{1,2}$, the
$(\x_0,\x_1,\x_2)$ space. We will demonstrate that in this loop formation
the parameters of Stokes matrices play an important role. 

For the four-gluon amplitude case specified by $\al = 0$ we will construct
a general string solution by combining two linearly independent solutions
with arbitrary two coefficients in the left and right linear problems
respectively.  The reality and normalization conditions for the 
general solutions constrain the left and right coefficients to be
characterized by two parameters. We will examine how the freedoms
expressed by two parameters are related with some conformal SO(2,4)
transformations. We will demonstrate that the string configuration
determined from the solutions of the linear problems indeed satisfies
the string equations of motion and the Virasoro constraints in the
embedding coordinates $(Y_{-1},Y_0,Y_1,Y_2)$ as well as the
$AdS_3$ global coordinates $(t,\rho,\phi)$. Using only the asymptotic form
of the exact general string solution we will extract the figure of a 
generic tetragonal Wilson loop composed of four null vectors in $R^{1,2}$.

\section{Asymptotic string solutions for the hexagonal and octagonal
Wilson loops}

We consider the approximate string solution with Euclidean worldsheet in
$AdS_3$ which gives the multi-gluon amplitude in planar $\mathcal{N}=4$
 SYM theory at strong coupling \cite{FAM}. 
Through a Pohlmeyer type reduction, the problem of string moving
$AdS_3$ whose worldsheet is paramerized by complex coordinates
$z, \bz$ is transformed to the auxiliary linear problem involving
a single field $\al(z,\bz)$ and a holomorphic polynomial $p(z)$ whose 
degree $n-2$ determines the number of the cusps of Wilson loop
to be $2n$. The field $\al(z,\bz)$ satisfies a generalized 
Sinh-Gordon equation 
\begin{equation}
\pa\bar{\pa}\al(z,\bz) - e^{2\al(z,\bz)} + 
|p(z)|^2e^{-2\al(z,\bz)} = 0
\label{sh}\end{equation}
so that the following SL(2) connections $B^L, B^R$ are flat
\begin{eqnarray}
B_z^L &=& \left( \begin{array}{cc} \frac{1}{2}\pa\al & - e^{\al}\\
- e^{-\al}p(z) & -\frac{1}{2}\pa\al \end{array}
\right),\; B_{\bz}^L = \left( 
\begin{array}{cc} - \frac{1}{2}\bar{\pa}\al & - e^{-\al}\bar{p}(\bz)
\\ - e^{\al} & \frac{1}{2}\bar{\pa}\al \end{array}\right),
\nonumber \\
B_z^R &=& \left( \begin{array}{cc} - \frac{1}{2}\pa\al & e^{-\al}p(z)\\
- e^{\al} & \frac{1}{2}\pa\al \end{array}\right), \;
B_{\bz}^R = \left( 
\begin{array}{cc} \frac{1}{2}\bar{\pa}\al &   -e^{\al}
\\  e^{-\al}\bar{p}(\bz) & -\frac{1}{2}\bar{\pa}\al \end{array}
\right).
\end{eqnarray}
For the connections $B^{L,R}$ the auxiliary left and right linear
problems are given by
\begin{eqnarray}
\pa \psi_{\al}^L + (B_z^L)_{\al}^{\,\beta}\psi_{\beta}^L = 0,
\hspace{1cm} 
\bar{\pa} \psi_{\al}^L + (B_{\bz}^L)_{\al}^{\,\beta}\psi_{\beta}^L = 0,
\nonumber \\
\pa \psi_{\dot{\al}}^R + (B_z^R)_{\dot{\al}}^{\,\dot{\beta}}
\psi_{\dot{\beta}}^R = 0, \hspace{1cm} 
\bar{\pa} \psi_{\dot{\al}}^R + (B_{\bz}^R)_{\dot{\al}}^{\,\dot{\beta}}
\psi_{\dot{\beta}}^R = 0, 
\label{al}\end{eqnarray}
whose two linearly independent solutions $\psi_{\al,a}^L, a=1,2$ and
$\psi_{\dot{\al},\dot{a}}^R, \dot{a}=1,2$ are normalized as 
\begin{equation}
\epsilon^{\beta \al}\psi_{\al,a}^L\psi_{\beta,b}^L = \epsilon_{ab},
\hspace{1cm}
\epsilon^{\dot{\beta} \dot{\al}}\psi_{\dot{\al},\dot{a}}^R
\psi_{\dot{\beta},\dot{b}}^R = \epsilon_{\dot{a}\dot{b}}.
\label{nc}\end{equation}

These left and right solutions lead to the spacetime configuration of
the string surface which is expressed as
\begin{equation}
Y_{a\dot{a}} = \left( \begin{array}{cc} Y_{-1} + Y_2 & Y_1 - Y_0 \\
Y_1 + Y_0 & Y_{-1} - Y_2 \end{array} \right)_{a,\dot{a}} =
\psi_{\al,a}^L M_1^{\al\dot{\beta}}\psi_{\dot{\beta},\dot{a}}^R,
\hspace{1cm} M_1^{\al\dot{\beta}} = \left( \begin{array}{cc} 1 & 0 \\
0 & 1 \end{array} \right),
\label{ym}\end{equation}
where the embedding coordinates $Y_{\mu}$ describe the $AdS_3$ space
by $-Y_{-1}^2 - Y_0^2 + Y_1^2 + Y_2^2 = 1$.
Introducing the complex coordinates $w$ and $\bar{w}$ by
$dw = \sqrt{p(z)}dz, d\bar{w} = \sqrt{\bar{p}(\bz)}d\bz$ the generalized
Sinh-Gordon equation (\ref{sh}) is simplified to be 
\begin{equation}
\pa_{w}\bar{\pa}_{\bar{w}}\hat{\al} - e^{2\hat{\al}} + e^{-2\hat{\al}} 
= 0, \hspace{1cm} \hat{\al} \equiv \al - \frac{1}{4}\log p\bar{p}.
\end{equation}
The left and right linear problems (\ref{al}) are also expressed by
complex variables $w, \bar{w}$ and the approximate left and
right solutions for large $w$  yield the following string
solution
\begin{equation}
Y_{a\dot{a}} = \frac{1}{\sqrt{2}}( c_a^{L,+}c_{\dot{a}}^{R,+}e^u
+ c_a^{L,-}c_{\dot{a}}^{R,-}e^{-u} - c_a^{L,-}c_{\dot{a}}^{R,+}e^v
+ c_a^{L,+}c_{\dot{a}}^{R,-}e^{-v} ),
\label{so}\end{equation}
where 
\begin{equation}
u = w + \bar{w} + \frac{w - \bar{w}}{i}, \hspace{1cm}
v = -(w + \bar{w}) + \frac{w - \bar{w}}{i}.
\end{equation}
The expression (\ref{so}) shows the asymptotic solution such that in each
quadrant of the $w$ plane only one of these terms dominates and
characterizes the spacetime string configuration near each cusp.
Owing to the Stokes phenomenon the coefficients in (\ref{so})
in the five Stokes sectors for the left problem are 
specified as \cite{FAM}
\begin{eqnarray}
[12]: c_{[12],a}^{L,+} &=& b_a^+, \hspace{2cm}  c_{[12],a}^{L,-} = b_a^-,
\nonumber \\
\,[23]: c_{[23],a}^{L,+} &=& b_a^+ + \ga_2^Lb_a^-,  \hspace{1cm}
c_{[23],a}^{L,-} = b_a^-, \nonumber \\
\;[34]: c_{[34],a}^{L,+} &=& b_a^+ + \ga_2^Lb_a^-, \hspace{1cm} 
c_{[34],a}^{L,-} = b_a^- + \ga_3^L( b_a^+ + \ga_2^Lb_a^- ), \nonumber \\
\,[45]: c_{[45],a}^{L,+} &=& b_a^+ + \ga_2^Lb_a^- + 
\ga_4^L[ b_a^- + \ga_3^L( b_a^+ + \ga_2^Lb_a^- ) ], \nonumber \\
   c_{[45],a}^{L,-} &=& b_a^- + \ga_3^L( b_a^+ + \ga_2^Lb_a^- ), 
\nonumber \\
\,[56]: c_{[56],a}^{L,+} &=& b_a^+ + \ga_2^Lb_a^- + 
\ga_4^L[ b_a^- + \ga_3^L( b_a^+ + \ga_2^Lb_a^- ) ], \nonumber \\
c_{[56],a}^{L,-} &=& b_a^- + \ga_3^L( b_a^+ + \ga_2^Lb_a^- )
+\ga_5^L[ b_a^+ + \ga_2^Lb_a^- + \ga_4^L( b_a^- + 
\ga_3^L( b_a^+ + \ga_2^Lb_a^- )) ], 
\label{cb}\end{eqnarray}
where $\ga_i^L$ is the Stokes parameter for the left problem and
the parameters $b_a^r \;(r=+,-)$
satisfy
\begin{equation}
b_a^+b_b^- - b_a^-b_b^+ = \epsilon_{ab}.
\label{bn}\end{equation}
The coefficients of the asymptotic solution for the right problem are
also expressed by the right Stokes parameter $\ga_i^R$ and
the corrseponding quantities with tildes $\tilde{b}_{\dot{a}}^r$ 
which also satisfy
\begin{equation}
\tilde{b}_{\dot{a}}^+\tilde{b}_{\dot{b}}^- - \tilde{b}_{\dot{a}}^-
\tilde{b}_{\dot{b}}^+ = \epsilon_{\dot{a}\dot{b}}.
\label{nb}\end{equation}
Using the minimal surface string solution in $AdS_3$ (\ref{so}) together
with (\ref{cb}) we describe the $AdS_3$ subspace of $AdS_4$ in terms of
the $AdS_4$ embedding coordinates with $Y_3=0$ and express the 
$AdS_4$ embedding of the surface as
\begin{equation}
\frac{1}{\tilde{r}} = Y_{-1}, \hspace{1cm} \x_0 = \frac{Y_0}{Y_{-1}},
\hspace{1cm} \x_{1,2} = \frac{Y_{1,2}}{Y_{-1}}.
\label{fp}\end{equation}

We consider the spacetime feature of the string configuration in the 
$AdS_4$ Poincare coordinates $(\tilde{r}, \x_0, \x_1, \x_2)$.
In the first quadrant at $Rew > 0, \; Imw > 0$ which belongs to [12]
Stokes sector for the left problem and [01] Stokes sector for the
right problem where $c_{[01],\dot{a}}^{R,+} = \tilde{b}_{\dot{a}}^+$,
the first term in (\ref{so}) dominates
so that we combine (\ref{ym}) with (\ref{so}) and (\ref{cb}) to obtain
\begin{equation}
\left( \begin{array}{cc} Y_{-1} + Y_2 & Y_1 - Y_0 \\
Y_1 + Y_0 & Y_{-1} - Y_2 \end{array} \right) = \frac{e^u}{\sqrt{2}}
\left( \begin{array}{cc} b_1^+\tilde{b}_1^+ & b_1^+\tilde{b}_2^+ \\
 b_2^+\tilde{b}_1^+ & b_2^+\tilde{b}_2^+ \end{array} \right).
\end{equation}
The asymptotic solution in the first cusp labelled by (1,1) is
expressed as
\begin{equation}
\frac{1}{\tilde{r}} = \frac{1}{2\sqrt{2}} X_+^1 e^u, \;\;
\x_0^1 = \frac{Y_-^1}{X_+^1}, \; \;\x_1^1 = \frac{Y_+^1}{X_+^1}, \;\;
\x_2^1 = \frac{X_-^1}{X_+^1}
\label{fi}\end{equation}
with $X_{\pm}^1 = b_1^+\tilde{b}_1^+ \pm b_2^+\tilde{b}_2^+, \;
Y_{\pm}^1 = b_2^+\tilde{b}_1^+ \pm b_1^+\tilde{b}_2^+$.
For large $u$ the string approaches the boundary of $AdS_4$
specified by $\tilde{r} = 0$. 
In the second quadrant at $Rew < 0, \; Imw > 0$ which belongs to 
both [12] Stokes sectors for the left and right problems 
 the third term in (\ref{so}) becomes a big term 
so that the asymptotic solution in the second cusp labelled by (2,1) is
given by
\begin{equation}
\frac{1}{\tilde{r}} = -\frac{1}{2\sqrt{2}} X_+^2 e^v, \;\;
\x_0^2 = \frac{Y_-^2}{X_+^2}, \;\; \x_1^2 = \frac{Y_+^2}{X_+^2}, \;\;
\x_2^2 = \frac{X_-^2}{X_+^2}
\label{se}\end{equation}
with $X_{\pm}^2 = b_1^-\tilde{b}_1^+ \pm b_2^-\tilde{b}_2^+, \;
Y_{\pm}^2 = b_2^-\tilde{b}_1^+ \pm b_1^-\tilde{b}_2^+$.
In the third quadrant at $Rew < 0, \; Imw < 0$ which belongs to [23]
Stokes sector for the left problem and [12] Stokes sector for
the right problem the second term in (\ref{so}) dominates and 
the string near the third (2,2) cusp is specified by
\begin{equation}
\frac{1}{\tilde{r}} = \frac{1}{2\sqrt{2}} X_+^3 e^{-u}, \;\;
\x_0^3 = \frac{Y_-^3}{X_+^3}, \;\; \x_1^3 = \frac{Y_+^3}{X_+^3}, \;\;
\x_2^3 = \frac{X_-^3}{X_+^3}
\label{th}\end{equation}
with $X_{\pm}^3 = b_1^-\tilde{b}_1^- \pm b_2^-\tilde{b}_2^-, \;
Y_{\pm}^3 = b_2^-\tilde{b}_1^- \pm b_1^-\tilde{b}_2^-$.
In the fourth quadrant at $Rew > 0, \; Imw < 0$ which belongs to 
both [23] Stokes sectors for the left and right problems
 the fourth term in (\ref{so}) dominates and 
the string near the fourth (3,2) cusp is specified by
\begin{equation}
\frac{1}{\tilde{r}} = \frac{1}{2\sqrt{2}} X_+^4 e^{-v}, \;\;
\x_0^4 = \frac{Y_-^4}{X_+^4}, \;\; \x_1^4 = \frac{Y_+^4}{X_+^4}, \;\;
\x_2^4 = \frac{X_-^4}{X_+^4}
\label{fo}\end{equation}
with 
\begin{equation}
X_{\pm}^4 = ( b_1^+ + \ga_2^Lb_1^- )\tilde{b}_1^- \pm 
( b_2^+ + \ga_2^Lb_2^- )\tilde{b}_2^-, \;
Y_{\pm}^4 = ( b_2^+ + \ga_2^Lb_2^- )\tilde{b}_1^-
 \pm ( b_1^+ + \ga_2^Lb_1^- )\tilde{b}_2^-.
\end{equation}
Succeedingly the solutions in the fifth (3,3) cusp, the sixth (4,3) cusp,
the seventh (4,4) cusp and the eighth (5,4) cusp are respectively  
described in order as
\begin{equation}
\frac{1}{\tilde{r}} = \frac{1}{2\sqrt{2}} X_+^k f_k(u,v), \;\;
\x_0^k = \frac{Y_-^k}{X_+^k}, \;\; \x_1^k = \frac{Y_+^k}{X_+^k}, \;\;
\x_2^k = \frac{X_-^k}{X_+^k}, \; k = 5, 6, 7, 8
\label{rx}\end{equation}
where 
\begin{eqnarray}
f_k(u,v) &=& (e^u,\; -e^v, \; e^{-u}, \; e^{-v}), \nonumber \\
X_{\pm}^k &=& B_1^k\tilde{B}_1^k \pm B_2^k\tilde{B}_2^k,\hspace{1cm}
 Y_{\pm}^k = B_2^k\tilde{B}_1^k \pm B_1^k\tilde{B}_2^k,
\nonumber \\
B_a^5 &=& b_a^+ + \ga_2^Lb_a^-, \hspace{2cm} \tilde{B}_{\dot{a}}^5 = 
\tilde{b}_{\dot{a}}^+ + \ga_2^R\tilde{b}_{\dot{a}}^-, \nonumber \\
B_a^6 &=& b_a^- + \ga_3^L( b_a^+ + \ga_2^Lb_a^- ),  \hspace{1cm}
\tilde{B}_{\dot{a}}^6 = \tilde{b}_{\dot{a}}^+ 
+ \ga_2^R\tilde{b}_{\dot{a}}^-,  \nonumber \\
B_a^7 &=& b_a^- + \ga_3^L( b_a^+ + \ga_2^Lb_a^- ),  \hspace{1cm}
\tilde{B}_{\dot{a}}^7 = \tilde{b}_{\dot{a}}^- + \ga_3^R( 
\tilde{b}_{\dot{a}}^+ + \ga_2^R\tilde{b}_{\dot{a}}^- ), \nonumber \\
B_a^8 &=& b_a^+ + \ga_2^Lb_a^-  + \ga_4^L( b_a^- + \ga_3^L( b_a^+
 + \ga_2^Lb_a^- )), \;
\tilde{B}_{\dot{a}}^8 = \tilde{b}_{\dot{a}}^- + \ga_3^R( 
\tilde{b}_{\dot{a}}^+ + \ga_2^R\tilde{b}_{\dot{a}}^- ).
\label{lx}\end{eqnarray}
From these expressions we can confirm that the vectors 
$\x_{\mu}^k -  \x_{\mu}^{k+1}$ are the null vectors
\begin{equation}
(\x_{\mu}^k -  \x_{\mu}^{k+1})^2 = 0, \hspace{1cm}
k =1,2, \cdots, 6
\end{equation}
where the normalization conditions of the parameters 
$b_a^r$ and $\tilde{b}_{\dot{a}}^r$ in (\ref{bn}) and (\ref{nb}) are used.

Here for the hexagonal Wilson loop with six cusps to be constructed in a
closed form at the $AdS_4$ boundary we require $\x_{\mu}^7 = \x_{\mu}^1$
which gives the following relations for the Stokes parameters of the
left and right problems
\begin{equation}
1 + \ga_2^L\ga_3^L = 0, \hspace{1cm} 1 + \ga_2^R\ga_3^R = 0.
\label{sp}\end{equation}
In ref. \cite{FAM} by analyzing the behavior of the approximate left
and right solutions when we go around once in the $z$ plane
( or $n/2$ times in the $w$ plane ), the following relations
for the Stokes matrices for both left and right problems are presented 
\begin{eqnarray}
S_p(\ga_1)S_n(\ga_2)S_p(\ga_3)\cdots S_p(\ga_n) = i(-1)^{\frac{n-3}{2}}
\sigma^2 \hspace{1cm} n \; \mathrm{odd}, \nonumber \\
S_p(\ga_1)S_n(\ga_2)\cdots S_n(\ga_n)e^{\sigma^3(w_s + \bar{w}_s)}
 = -(-1)^{n/2} \hspace{1cm} n \; \mathrm{even}
\label{ss}\end{eqnarray}
with the Pauli matrices $\sigma^i$, where
\begin{equation}
S_p = \left( \begin{array}{cc} 1 & \ga \\ 0 & 1 \end{array} \right),
\hspace{1cm} 
S_n = \left( \begin{array}{cc} 1 & 0 \\ \ga & 1 \end{array} \right)
\end{equation}
and the shift parameters $w_s, \bar{w}_s$ are related with the
spacetime cross ratios. For the hexagonal Wilson loop case
we express the first relation in (\ref{ss}) with $n = 3$ as
\begin{equation}
\left( \begin{array}{cc} 1 + \ga_1\ga_2 & \ga_1 +\ga_3 + \ga_1\ga_2\ga_3
\\ \ga_2 & 1 + \ga_2\ga_3 \end{array} \right) =
\left( \begin{array}{cc} 0 & 1 \\ -1 & 0 \end{array} \right),
\label{sr}\end{equation}
whose diagonal components directly yield (\ref{sp}).
Alternatively this observation implies that the segment between the first
cusp and the sixth cusp is described by a lightlike vector.

In order to consider the $n = 4$ case, that is, the octagonal Wilson loop
case we write down the solution in the nineth (5,5) cusp as 
the forms of (\ref{rx}) and (\ref{lx}) 
for $k = 9$ with $f_9 = e^u$ and 
\begin{eqnarray}
B_a^9 &=& b_a^+ + \ga_2^Lb_a^-  + \ga_4^L( b_a^- + \ga_3^L( b_a^+
 + \ga_2^Lb_a^- )), \nonumber \\
\tilde{B}_{\dot{a}}^9 &=& \tilde{b}_{\dot{a}}^+ 
+ \ga_2^R\tilde{b}_{\dot{a}}^-  
+ \ga_4^R( \tilde{b}_{\dot{a}}^- + \ga_3^R( \tilde{b}_{\dot{a}}^+ 
+ \ga_2^R\tilde{b}_{\dot{a}}^- )).
\end{eqnarray}
For the eight-cusp Wilson loop to be of a closed form we require
$\x_{\mu}^9 = \x_{\mu}^1$ which leads to the constraints for the
Stokes parameters
\begin{equation}
\ga_2^L + \ga_4^L + \ga_2^L\ga_3^L\ga_4^L = 0, \hspace{1cm} 
\ga_2^R + \ga_4^R + \ga_2^R\ga_3^R\ga_4^R = 0.
\label{rl}\end{equation}
The second relation in (\ref{ss}) with $n = 4$ is given by
\begin{equation}
\left( \begin{array}{cc} [(1 + \ga_2\ga_3)(1 + \ga_3\ga_4) + \ga_1\ga_4]
e^{w_s + \bar{w}_s}  & (\ga_1 + \ga_3 + \ga_1\ga_2\ga_3)
e^{-(w_s + \bar{w}_s)}   \\
(\ga_2 + \ga_4 + \ga_2\ga_3\ga_4)e^{w_s + \bar{w}_s} &  
(1 + \ga_2\ga_3)e^{-(w_s + \bar{w}_s)}  
\end{array} \right) = -\left( \begin{array}{cc} 1 & 0 \\ 0 & 1 \end{array}
\right),
\end{equation}
whose off-diagonal components directly yield (\ref{rl}).

For the hexagonal Wilson loop case in the projected $(\x_1, \x_2)$ plane
the square of the position vector $\tilde{\vx}^1 = (\x_1^1,\x_2^1)$
 for the first cusp is larger than unity
\begin{equation}
(\tilde{\vx}^1)^2  = \frac{\vec{b}_L^2 \vec{b}_R^2}
{(\vec{b}_L \cdot \vec{b}_R)^2} \ge 1,
\label{fr}\end{equation}
where two vectors  $\vec{b}_L$ and $\vec{b}_R$ are defined by
$\vec{b}_L = ( b_1^+, b_2^+ ), \; \vec{b}_R = 
( \tilde{b}_1^+, \tilde{b}_2^+ )$.
In similar expressions including an inner product of two vectors
we see $(\tilde{\vx}^k)^2 \ge 1, \; k = 1, 2, \cdots, 6$. For example 
the sixth cusp $(k = 6)$ is  characterized by
\begin{equation}
\vec{b}_L = (B_1^6, \; B_2^6), \hspace{1cm} \vec{b}_R = 
(\tilde{B}_1^6, \; \tilde{B}_2^6).
\end{equation}
Thus we observe that in the $(\x_1, \x_2)$ plane each cusp is located
outside the unit circle.

An arbitrary point $P_k \; (k = 1,2, \cdots, 6)$ between $\tilde{\vx}^k$ 
and $\tilde{\vx}^{k+1}$ defines a vector
\begin{equation}
\mbox{\boldmath{$P$}}_k = \tilde{\vx}^k + ( \tilde{\vx}^{k+1} 
- \tilde{\vx}^k)t_k
\label{tk}\end{equation}
with $\tilde{\vx}^7 = \tilde{\vx}^1$,
whose parameter $t_k$ is fixed as
\begin{equation}
t_k = \frac{(\tilde{\vx}^k)^2 - \tilde{\vx}^k\cdot\tilde{\vx}^{k+1}}
{(\tilde{\vx}^k - \tilde{\vx}^{k+1})^2}
\label{tx}\end{equation}
by demanding the orthogonal condition 
$\mbox{\boldmath{$P$}}_k \cdot (\tilde{\vx}^k - \tilde{\vx}^{k+1}) = 0$.
Therefore eliminating $t_k$ we have 
\begin{equation}
\mbox{\boldmath{$P$}}_k^2 = \frac{(\tilde{\vx}^k )^2 (\tilde{\vx}^{k+1})^2
 - (\tilde{\vx}^k \cdot  \tilde{\vx}^{k+1})^2}
{(\tilde{\vx}^k - \tilde{\vx}^{k+1})^2}, 
\label{kp}\end{equation}
which can be shown to become unity through (\ref{bn}) and (\ref{nb}),
 so that each point $P_k$ is located on the unit circle. 
Thus we demonstrate that
for $n = 3$ the sides of hexagonal Wilson loop are tangent to the
unit circle in the projected  $(\x_1, \x_2)$ plane.
In the $(\x_0, \x_1, \x_2)$ space the time component of the $k$-th
point $P_k$ is defined by $\x_0^k + (\x_0^{k+1}-  \x_0^k)t_k$
which turns out to be zero through (\ref{tx}). Thus in the 
$(\x_0, \x_1, \x_2)$ space the cusps of hexagonal Wilson loop are going
alternating up and down outside the unit circle. It implies that the
even sides of the Wilson loop allow the successive lightlike vectors to
form a closed contour.

Let us examine how the asymptotic solution (\ref{so}) constructed from the
auxiliary linear problems in the $w$ plane
satisfies the string equations of motion and the
Virasoro constraints in the $z$ plane for the octagonal Wilson loop
case $(n = 4)$. For large $z$ or large $w$, the polynomials 
$p(z),\; \bar{p}(\bz)$ are approximately given by   
$p(z) \approx z^2,\; \bar{p}(\bar{z}) \approx \bz^2$ so that 
$w \approx z^2/2,\; \bar{w} \approx \bar{z}^2/2$. We use a notation
$z = x + iy, \;  \bz = x - iy$ to express $u, v$ as
$u \approx x^2 - y^2 + 2xy, \; v \approx - x^2 + y^2 + 2xy$.
The conformal gauge equations of motion and the Virasoro constraints in
the $z$ plane are expressed in terms of the embedding coordinates
$Y_{\mu}$ as
\begin{equation}
\pa \bar{\pa}Y_{\mu} - (\pa Y_{\nu}\bar{\pa}Y^{\nu})Y_{\mu} = 0,
\hspace{1cm} \pa Y_{\mu}\pa Y^{\mu} = \bar{\pa}Y_{\mu}\bar{\pa} Y^{\mu}
= 0.
\label{ev}\end{equation}
Here we begin with the asymptotic solutions in the first, second,
third and fourth quadrants
\begin{equation}
(Y_{-1}^k, Y_0^k, Y_1^k, Y_2^k) = \frac{1}{2\sqrt{2}}( X_+^k, Y_-^k, 
 Y_+^k,  X_-^k)f_k(x,y), \;\; k = 1, 2, 3, 4
\label{xy}\end{equation}
where
\begin{equation}
f_1 = e^{ x^2 - y^2 + 2xy}, \; f_2 = -e^{- x^2 + y^2 + 2xy}, \;
f_3 = e^{- x^2 + y^2 - 2xy},\; f_4 = e^{ x^2 - y^2 - 2xy}.
\label{fk}\end{equation}
For large $w$ the parameter $\hat{\al}$ becomes zero so that 
$\pa Y_{\nu}\bar{\pa}Y^{\nu} = 2e^{2\al} \approx 2\sqrt{p\bar{p}}
\approx 2(x^2 + y^2)$. This is compared with the four-side Wilson loop
case $(n = 2)$ where $\hat{\al} = \al = 0$ and the $w$ plane is
identical to the $z$ plane. The equations of motion for large $z$
\begin{equation}
\frac{1}{4}(\pa_x^2 + \pa_y^2)Y_{\mu}^k - 2(x^2 + y^2)Y_{\mu}^k = 0
\label{ez}\end{equation}
are satisfied by (\ref{xy}) with 
the four kinds of functions $f_k(x,y)$ in (\ref{fk}).
The Virasoro constraints also hold owing to the relations 
$(X_+^k)^2 - (X_-^k)^2 = (Y_+^k)^2 - (Y_-^k)^2, k = 1, 2, 3, 4$
which follow from (\ref{fi}), (\ref{se}), (\ref{th}) and (\ref{fo}).
For the other quadrant each dominant solution in (\ref{so}) for large
$z$ satisfies the equations of motion and the Virasoro constraints
through the same relations $(X_+^k)^2 - (X_-^k)^2 = (Y_+^k)^2 
- (Y_-^k)^2, k = 5,6,7,8$ which are obeyed by the expressions
in (\ref{lx}). These relations are attributed to the expression that
the coefficient of the dominant term for each quadrant in (\ref{so})
is given in a product form as $c_a^{L\pm}c_{\dot{a}}^{R\pm}$.

\section{Asymptotic behaviors of the general string solutions for 
the tetragonal Wilson loop}

We consider the four-side Wilson loop in the $z$ plane which is specified
by $\hat{\al} = \al = 0, p = 1$. The left linear problem in (\ref{al})
becomes
\begin{eqnarray}
\pa_z \psi_1^L &=&  \psi_2^L,  \hspace{1cm} \pa_z \psi_2^L = \psi_1^L, 
\label{lf} \\
\bar{\pa}_{\bz} \psi_1^L &=& \psi_2^L,   \hspace{1cm} 
\bar{\pa}_{\bz}\psi_2^L = \psi_1^L.
\label{ls}\end{eqnarray}
The two equations in (\ref{lf}) combine to be 
$\pa_z^2 \psi_1^L = \psi_1^L$ which gives two linearly independent
solutions $\psi_{1a}^L, \; a = 1, 2$ as expressed by
$\psi_{11}^L = c_1(\bar{z})e^z, \; \psi_{12}^L = c_2(\bar{z})e^{-z}$.
The two solutions should satisfy $\bar{\pa}_{\bz}^2 \psi_1^L = \psi_1^L$
which follows from (\ref{ls}) so that $c_1(\bz), c_2(\bz)$ are
expanded by $c_1(\bz) = c_{11}e^{\bz} + c_{12}e^{-\bz}, \;
c_2(\bz) = c_{21}e^{\bz} + c_{22}e^{-\bz}$.
The substitution of $\psi_{1a}^L$ into (\ref{lf}) gives
$\psi_{21}^L = c_1(\bz)e^z, \; \psi_{22}^L = - c_2(\bz)e^{-z}$.
The expressions $\psi_{11}^L$ and $\psi_{21}^L$ are substituted into 
(\ref{ls}) and we have $c_{12} = 0$, while the insertions of 
$\psi_{12}^L$ and $\psi_{22}^L$ into (\ref{ls}) lead to $c_{21} = 0$.
Thus we obtain two linearly independent solutions for the left
problem
\begin{equation}
\psi_{\al 1}^L = c_{11}\left( \begin{array}{c} e^{z + \bz} \\
e^{z + \bz} \end{array} \right), \hspace{1cm}
\psi_{\al 2}^L = c_{22}\left( \begin{array}{c} e^{- z - \bz} \\
- e^{- z - \bz} \end{array} \right),
\end{equation}
whose coefficients are fixed as $c_{11} = c_{22} = 1/\sqrt{2}$ through
the normalization condition (\ref{nc}).

The right linear problem in (\ref{al}) is also expressed as
\begin{equation}
\pa_z \psi_1^R = - \psi_2^R, \; \pa_z \psi_2^R = \psi_1^R, \;
\bar{\pa}_{\bz} \psi_1^R = \psi_2^R, \; 
\bar{\pa}_{\bz} \psi_2^R = -\psi_1^R,
\end{equation}
which give two linearly independent solutions
$\psi_{11}^R = \tilde{c}_1(\bz)e^{-iz}, \; \psi_{12}^R = 
\tilde{c}_2(\bz)e^{iz}$ and $\psi_{21}^R = i\tilde{c}_1(\bz)
e^{-iz}, \; \psi_{22}^R = - i\tilde{c}_2(\bz)e^{iz}$ where
the coefficients are expanded as 
$\tilde{c}_1(\bz) = \tilde{c}_{11}e^{i\bz} + \tilde{c}_{12}
e^{-i\bz}$ and $\tilde{c}_2(\bz) = \tilde{c}_{21}e^{i\bz}
+ \tilde{c}_{22}e^{-i\bz}$ but with $\tilde{c}_{12} = 
\tilde{c}_{21} = 0$. The two linearly independent solutions for the
right problem are expressed as
\begin{equation}
\psi_{\al 1}^R = \tilde{c}_{11}\left( \begin{array}{c} 
e^{\frac{z - \bz}{i}} \\ ie^{\frac{z - \bz}{i}} \end{array}
 \right), \hspace{1cm}
\psi_{\al 2}^R = \tilde{c}_{22}\left( \begin{array}{c} 
e^{-\frac{z - \bz}{i}} \\ - ie^{-\frac{z - \bz}{i}} 
\end{array} \right),
\end{equation}  
whose coefficients are fixed as $\tilde{c}_{11} = 1/\sqrt{2}, \;
\tilde{c}_{22} = -i/\sqrt{2}$ through the normalization condition 
(\ref{nc}). 

For the left and right problems the general solutions are given by
the linear combination of two independent solutions
\begin{equation}
{\psi'}_{\al a}^L = d_{ab}\psi_{\al b}^L, \hspace{1cm}
{\psi'}_{\al \dot{a}}^R = \tilde{d}_{\dot{a}\dot{b}}
\psi_{\al \dot{b}}^R.
\label{gd}\end{equation}
If we substitute these expressions into the normalization condition
(\ref{nc}) for $\psi'$ we have
\begin{equation}
d_{11}d_{22} - d_{12}d_{21} = 1, \hspace{1cm}  \tilde{d}_{11}
\tilde{d}_{22}  - \tilde{d}_{12}\tilde{d}_{21} = 1.
\label{dd}\end{equation}
Substitutions of the general solutions (\ref{gd}) into (\ref{ym})
yield
\begin{eqnarray}
Y_{-1} + Y_2 &=& \frac{1}{2}[d_{11}(1 + i)(\tilde{d}_{11}e^u
-  \tilde{d}_{12}e^{-v} ) + d_{12}(1 - i)(\tilde{d}_{11}e^v
+  \tilde{d}_{12}e^{-u} ) ], \nonumber \\
Y_{-1} - Y_2 &=& \frac{1}{2}[d_{21}(1 + i)(\tilde{d}_{21}e^u
-  \tilde{d}_{22}e^{-v} ) + d_{22}(1 - i)(\tilde{d}_{21}e^v
+  \tilde{d}_{22}e^{-u} ) ], \nonumber \\
Y_1 + Y_0 &=& \frac{1}{2}[d_{21}(1 + i)(\tilde{d}_{11}e^u
-  \tilde{d}_{12}e^{-v} ) + d_{22}(1 - i)(\tilde{d}_{11}e^v
+  \tilde{d}_{12}e^{-u} ) ], \nonumber \\
Y_1 - Y_0 &=& \frac{1}{2}[d_{11}(1 + i)(\tilde{d}_{21}e^u
-  \tilde{d}_{22}e^{-v} ) + d_{12}(1 - i)(\tilde{d}_{21}e^v
+  \tilde{d}_{22}e^{-u} ) ],
\end{eqnarray}
where $u = z + \bz + (z - \bz)/i, \; v = - (z + \bz)
 + (z - \bz)/i$. It is noted that we should choose the parameters
$d_{ab}$ such that $d_{11}, d_{21}$ include a factor $(1 - i)$
and $d_{22}, d_{12}$ have $(1 + i)$ because $Y_{\mu}$ is real.
Therefore we use the conditions in (\ref{dd}) to parametrize
$d_{ab}, \tilde{d}_{\dot{a}\dot{b}}$ as
\begin{equation}
d_{ab} = \frac{1}{\sqrt{2}}\left( \begin{array}{cc}
(1 - i)\cos\beta & (1 + i)\sin\beta \\ -(1 - i)\sin\beta
& (1 + i)\cos\beta \end{array} \right), \;
\tilde{d}_{\dot{a}\dot{b}} = \left( \begin{array}{cc}\cosh\ga & \sinh\ga
 \\ \sinh\ga & \cosh\ga \end{array} \right). 
\end{equation}
Here if we express the complex variables $z, \bz$ as $z = (\sigma +
i\tau )/2, \; \bz = (\sigma - i\tau )/2$ the string configuration
is obtained by
\begin{eqnarray}
\left( \begin{array}{c} Y_{-1} \\ Y_1 \end{array} \right) &=&
\frac{1}{\sqrt{2}}\left( \begin{array}{cc}\cosh\ga & \sinh\ga \\ 
\sinh\ga & \cosh\ga \end{array} \right)
\left( \begin{array}{c}
\cos\beta\cosh(\sigma + \tau) + \sin\beta\cosh(\sigma - \tau)  \\
-\sin\beta\sinh(\sigma + \tau) - \cos\beta\sinh(\sigma - \tau)
 \end{array} \right), \label{ge} \\
\left( \begin{array}{c} Y_0 \\ Y_2 \end{array} \right) &=&
\frac{1}{\sqrt{2}}\left( \begin{array}{cc}\cosh\ga & -\sinh\ga \\ 
-\sinh\ga & \cosh\ga \end{array} \right)
\left( \begin{array}{c}
-\sin\beta\cosh(\sigma + \tau) + \cos\beta\cosh(\sigma - \tau)  \\
\cos\beta\sinh(\sigma + \tau) - \sin\beta\sinh(\sigma - \tau) 
\end{array} \right). \nonumber
\end{eqnarray}

For the $\ga = 0, \beta \ne 0$ case the string solution is expressed as
\begin{eqnarray}
\left( \begin{array}{c} Y_{-1} \\ Y_0 \end{array} \right) &=&
\left( \begin{array}{cc}\cos\beta & \sin\beta \\ 
-\sin\beta & \cos\beta \end{array} \right)
\left( \begin{array}{c}\frac{1}{\sqrt{2}}\cosh(\sigma + \tau) \\
\frac{1}{\sqrt{2}}\cosh(\sigma - \tau) \end{array}\right), 
\nonumber \\
\left( \begin{array}{c} Y_1 \\ Y_2 \end{array} \right) &=&
\left( \begin{array}{cc}\cos\beta & -\sin\beta \\ 
\sin\beta & \cos\beta \end{array} \right)
\left( \begin{array}{c} -\frac{1}{\sqrt{2}}\sinh(\sigma - \tau) \\
\frac{1}{\sqrt{2}}\sinh(\sigma + \tau)\end{array} \right).
\label{sb}\end{eqnarray}
The basic solution specified by $\ga = 0, \;\beta =\pi/4$ is given by
\begin{equation}
Y_{-1} = \cosh\tau\cosh\sigma, \; Y_0 = -\sinh\tau\sinh\sigma, \;
Y_1 = -\cosh\tau\sinh\sigma, \; Y_2 = \sinh\tau\cosh\sigma.
\label{ba}\end{equation}
The sign change as $\sigma \rightarrow -\sigma$ for (\ref{ba})
leads to the expression in ref. \cite{AM} associated with the 
square Wison loop with four cusps which is obtained by making 
a scaling limit of the spinning folded string 
with Minkowski worldsheet in $AdS_3$ and changing to Euclidean 
worldsheet \cite{KRT}. On the other hand the $\ga = \beta = 0$ solution
is written by
\begin{eqnarray}
Y_{-1} = \frac{1}{\sqrt{2}}\cosh(\sigma + \tau), \hspace{1cm}
 Y_0 = \frac{1}{\sqrt{2}}\cosh(\sigma - \tau), \nonumber \\
Y_1 = -\frac{1}{\sqrt{2}}\sinh(\sigma - \tau), \hspace{1cm} 
Y_2 = \frac{1}{\sqrt{2}}\sinh(\sigma + \tau),
\label{yo}\end{eqnarray}
which is transformed through $\sigma \rightarrow -\sigma$ and 
a discrete SO(2,4) interchange 
$Y_{-1}\leftrightarrow Y_0$ to the solution \cite{KRT} which is 
obtained by making an SO(2) rotation in the $(\tau, \sigma)$ plane for the
basic one-cusp solution \cite{MK}. Thus we have observed that the linear
coefficients $d_{ab}$ for the left problem are parametrized to be 
associated with the simultaneous SO(2) rotations with opposite angles in 
the $(Y_{-1}, Y_0)$ and $(Y_1, Y_2)$  planes, while those 
$\tilde{d}_{\dot{a}\dot{b}}$
for the right problem with the simultaneous SO(2,4) boosts 
with opposite boost parameters in the $(Y_{-1}, Y_1)$ and $(Y_0, Y_2)$
planes. As the other parametrizations we take $\beta = \pi/4$
with
\begin{equation}
\tilde{d}_{\dot{a}\dot{b}} = \left( \begin{array}{cc}e^{\tau_1} & 0 \\ 
0 & e^{-\tau_1} \end{array} \right), \hspace{1cm}
\tilde{d}_{\dot{a}\dot{b}} = \left( \begin{array}{cc} 0 & e^{\tau_2} \\ 
 -e^{-\tau_2} & 0\end{array} \right)
\end{equation}
to derive the string solutions respectively as
\begin{eqnarray}
Y_{-1} &=& \cosh(\tau + \tau_1)\cosh\sigma, \hspace{1cm}
Y_0  = -\sinh(\tau + \tau_1)\sinh\sigma, \nonumber \\
Y_1 &=& -\cosh(\tau + \tau_1)\sinh\sigma, \hspace{1cm}
Y_2  = \sinh(\tau + \tau_1)\cosh\sigma
\label{ta}\end{eqnarray}
and
\begin{eqnarray}
Y_{-1} &=& \sinh(\tau - \tau_2)\sinh\sigma, \hspace{1cm}
Y_0  = \cosh(\tau - \tau_2)\cosh\sigma, \nonumber \\
Y_1 &=& -\sinh(\tau - \tau_2)\cosh\sigma, \hspace{1cm}
Y_2  = -\cosh(\tau - \tau_2)\sinh\sigma.
\end{eqnarray}
The configuration (\ref{ta}) shows only a shift of $\tau$ by $\tau_1$
for the basic solution (\ref{ba}). 

The Virasoro constraints in (\ref{ev}) are given by 
\begin{equation}
( \pa Y_{-1} \; \pa Y_1 )
\left( \begin{array}{cc}-1 & 0 \\ 0 & 1 \end{array}\right)
\left( \begin{array}{c}\pa Y_{-1} \\ \pa Y_1 \end{array}\right) +
( \pa Y_0 \; \pa Y_2 )
\left( \begin{array}{cc}-1 & 0 \\ 0 & 1 \end{array}\right)
\left( \begin{array}{c}\pa Y_0 \\ \pa Y_2 \end{array}\right) = 0
\end{equation}
and the equation which we get by the replacement of 
$\pa \rightarrow \bar{\pa}$, which are 
confirmed to be satisfied by the matrix representation of the
general solution (\ref{ge}).  Similarly using 
the matrix representation (\ref{ge}) we obtain  
$\pa Y_{\nu} \bar{\pa}Y^{\nu} = 2$, that is, $\al = 0$ consistently so 
that the equations of motion in (\ref{ev}) become 
\begin{equation}
(\pa_{\sigma}^2 + \pa_{\tau}^2)Y_{\mu} - 2Y_{\mu}= 0 \hspace{0.5cm}
\mathrm{or} \hspace{0.5cm} \frac{1}{4}(\pa_x^2 + \pa_y^2)Y_{\mu} 
- 2Y_{\mu}= 0, 
\end{equation}
which are simply satisfied by the string configuration (\ref{ge}) and
 compared with the asymptotic equations of motion (\ref{ez}) for large 
$z$ associated with the string configuration for the hexagonal Wilson
loop case.

The embedding coordinates $Y_{\mu}$ are related to the standard global 
coordinates $(t, \rho, \phi)$ on $AdS_3$ by
\begin{equation}
Y_{-1} + iY_0 = \cosh\rho e^{it}, \hspace{1cm} Y_1 + iY_2 = \sinh\rho 
e^{i\phi}.
\end{equation}
We analayze the general solution (\ref{ge}) in the global 
coordinates, which is expressed as
\begin{eqnarray}
&\cosh\rho = \frac{1}{\sqrt{2}}[ (\cosh\ga \cosh\sigma_+ - 
\sinh\ga \sinh\sigma_-)^2 + (\sinh\ga \sinh\sigma_+ - 
\cosh\ga \cosh\sigma_-)^2 ]^{1/2}& \nonumber \\
& = [ \cosh^2\tau \cosh^2(\sigma - \ga) + \sinh^2\tau 
\sinh^2(\sigma + \ga)]^{1/2},& \nonumber \\
&\sinh\rho = \frac{1}{\sqrt{2}}[ (\sinh\ga \cosh\sigma_+ - 
\cosh\ga \sinh\sigma_-)^2 + (\cosh\ga \sinh\sigma_+ - 
\sinh\ga \cosh\sigma_-)^2 ]^{1/2}& \nonumber \\
 &= [ \cosh^2\tau \sinh^2(\sigma - \ga) + \sinh^2\tau 
\cosh^2(\sigma + \ga)]^{1/2},& \nonumber \\ 
&\tan(t + \beta) = \frac{-\sinh\ga \sinh\sigma_+ + 
\cosh\ga \cosh\sigma_-}{\cosh\ga \cosh\sigma_+ - \sinh\ga \sinh\sigma_-},&
\nonumber \\
&\tan(\phi - \beta) = \frac{ \cosh\ga \sinh\sigma_+ - 
\sinh\ga \cosh\sigma_-}{\sinh\ga \cosh\sigma_+ - \cosh\ga \sinh\sigma_-}&
\label{rt}\end{eqnarray}
with $\sigma_{\pm} = \sigma \pm \tau$.
Since there are compact differential expressions derived from (\ref{rt})
\begin{eqnarray}
\pa_{\tau}t &=& - \frac{\sinh(\sigma + \ga)\cosh(\sigma - \ga)}
{\cosh^2\rho}, \hspace{1cm}
\pa_{\sigma}t = - \frac{\cosh2\ga \sinh 2\tau}{2\cosh^2\rho},
\nonumber \\
\pa_{\tau}\phi &=& - \frac{\sinh(\sigma - \ga)\cosh(\sigma + \ga)}
{\sinh^2\rho}, \hspace{1cm}
\pa_{\sigma}\phi = \frac{\cosh2\ga \sinh 2\tau}{2\sinh^2\rho},
\label{de}\end{eqnarray}
the equations of motion for $t$ and $\phi$ from the conformal gauge string
Lagrangian given by
\begin{eqnarray}
\pa_{\tau}(\cosh^2\rho\pa_{\tau}t) + \pa_{\sigma}(\cosh^2\rho
\pa_{\sigma}t) = 0, \nonumber \\
\pa_{\tau}(\sinh^2\rho\pa_{\tau}\phi) + \pa_{\sigma}(\sinh^2\rho
\pa_{\sigma}\phi) = 0
\end{eqnarray}
are simply satisfied. The other symmetric differential expressions
\begin{equation}
\pa_{\tau}\rho = \frac{\cosh 2\ga\sinh 2\tau\cosh 2\sigma}{\sinh 2\rho},\;
\pa_{\sigma}\rho = \frac{\cosh^2\tau\sinh 2(\sigma - \ga) +
\sinh^2\tau \sinh 2(\sigma + \ga)}{\sinh 2\rho}
\label{rr}\end{equation}
with (\ref{de}) make the Virasoro constraint $T_{\tau\sigma}= 0$,
\begin{equation}
\pa_{\tau}\rho\pa_{\sigma}\rho = \cosh^2\rho\pa_{\tau}t\pa_{\sigma}t
- \sinh^2\rho\pa_{\tau}\phi \pa_{\sigma}\phi
\end{equation}
hold through the suitable variables $\sigma \pm \ga$.
The other Virasoro constraint $T_{\tau\tau} - T_{\sigma\sigma} = 0$
for the Euclidean worldsheet described by
\begin{equation}
(\pa_{\tau}\rho)^2 - (\pa_{\sigma}\rho)^2 = \cosh^2\rho((\pa_{\tau}t)^2
- (\pa_{\sigma}t)^2) - \sinh^2\rho((\pa_{\tau}\phi)^2 - 
(\pa_{\sigma}\phi)^2 ) 
\end{equation}
is also satisfied. The equation of motion for $\rho$
\begin{equation}
\pa_{\tau}^2\rho + \pa_{\sigma}^2\rho = \frac{1}{2}\sinh2\rho ( 
(\pa_{\tau}\phi)^2 + (\pa_{\sigma}\phi)^2 - (\pa_{\tau}t)^2 -
(\pa_{\sigma}t)^2 )
\end{equation}
can be confirmed to be satisfied by using $\cosh^2(A \pm B) + 
\sinh^2(A \mp B) = \cosh2A \cosh2B$ and the $\cosh2\tau$
expression of $\pa_{\sigma}\rho$ in (\ref{rr}).  

The general string solution (\ref{ge}) is expressed in terms of the 
$AdS_4$ Poincare coordinates (\ref{fp}) as
\begin{eqnarray}
\frac{1}{\tilde{r}} = \frac{1}{\sqrt{2}}[ \cosh\ga(\cos\beta \cosh\sigma_+
+ \sin\beta \cosh\sigma_-) -  \sinh\ga(\sin\beta \sinh\sigma_+
+ \cos\beta \sinh\sigma_-) ], \nonumber \\
\x_0 = \frac{\tilde{r}}{\sqrt{2}}[ \cosh\ga(-\sin\beta \cosh\sigma_+
+ \cos\beta \cosh\sigma_-) -  \sinh\ga(\cos\beta \sinh\sigma_+
- \sin\beta \sinh\sigma_-) ], \nonumber \\
\x_1 = \frac{\tilde{r}}{\sqrt{2}}[ \sinh\ga(\cos\beta \cosh\sigma_+
+ \sin\beta \cosh\sigma_-) -  \cosh\ga(\sin\beta \sinh\sigma_+
+ \cos\beta \sinh\sigma_-) ], \nonumber \\
\x_2 = \frac{\tilde{r}}{\sqrt{2}}[ \sinh\ga(\sin\beta \cosh\sigma_+
- \cos\beta \cosh\sigma_-) +  \cosh\ga(\cos\beta \sinh\sigma_+
- \sin\beta \sinh\sigma_-) ].
\label{po}\end{eqnarray}
For the basic $\ga = 0, \; \beta = \pi/4$ solution we have
\begin{equation}
\tilde{r} = \frac{1}{\cosh\tau \cosh\sigma}, \; \x_0 = -\tanh\tau
\tanh\sigma, \; \x_1 = - \tanh\sigma, \; \x_2 = \tanh\tau,
\label{rf}\end{equation}
where the eliminations of $\sigma, \tau$ lead to 
$\tilde{r}= \sqrt{(1-\x_1^2)(1-\x_2^2)}, \; \x_0 = \x_1\x_2$ whose 
expressions directly give the positions of the four cusps for the
square Wilson loop. For the $\ga = \beta = 0$ solution we have
\begin{equation}
\tilde{r} = \frac{\sqrt{2}}{\cosh\sigma_+},\; \x_0 = \frac{\cosh\sigma_-}
{\cosh\sigma_+}, \; \x_1 = -\frac{\sinh\sigma_-}{\cosh\sigma_+}, \; 
 \x_2 = \tanh\sigma_+.
\end{equation}
The eliminations of variables $\sigma_+$ and  $\sigma_-$ yield
\begin{equation}
\tilde{r} = \sqrt{2(1 - \x_2^2)}, \hspace{1cm}  \x_0 = 
\sqrt{ 1 + \x_1^2 - \x_2^2 },
\label{tc}\end{equation}
which provide the locations of two cusps as $(\x_0, \x_1, \x_2) =
(0, 0, 1), \; (0, 0, -1)$ at which two semi infinite lightlike lines
intersect transversely (see the first ref. in \cite{SR}).

On the other hand we cannot eliminate the $\sigma_+, \sigma_-$
variables for the general solution (\ref{po}) but 
it is possible to extract the
cusp positions in the $(\x_0, \x_1, \x_2)$ space by analyzing
the asymptotic behavior of (\ref{po}). The $AdS_4$ boundary is 
chracterized by $\tilde{r}= 0$, which is achieved by taking the four
limits as $\sigma_+ \rightarrow \pm \infty$ and  $\sigma_- \rightarrow \pm
\infty$. Defining the variables $y_i, i = 1, 2,\cdots, 6$ as
\begin{eqnarray}
y_1 &=& \frac{\tan\beta - \tanh\ga}{1 - \tan\beta\tanh\ga}, \hspace{0.5cm}
y_2 = \frac{\tan\beta + \tanh\ga}{1 + \tan\beta\tanh\ga},   \hspace{0.5cm}
y_3 = \frac{1 + \tan\beta\tanh\ga}{1 - \tan\beta\tanh\ga},
\nonumber \\
y_4 &=& \frac{\tan\beta - \tanh\ga}{\tan\beta + \tanh\ga}, \hspace{0.5cm}
y_5 = \frac{1 - \tan\beta\tanh\ga}{\tan\beta + \tanh\ga},   \hspace{0.5cm}
y_6 = \frac{1 + \tan\beta\tanh\ga}{\tan\beta - \tanh\ga},
\label{yv}\end{eqnarray}
we express the first cusp obtained by taking $\sigma_+ \rightarrow 
\infty$ limit as $\x_{\mu}^1 = (-1/y_5, -y_1, y_3)$, the second one 
in the $\sigma_- \rightarrow -\infty$ limit as  $\x_{\mu}^2 = (y_5, 
1/y_2, y_4)$, the third one in the $\sigma_+ \rightarrow -\infty$ limit 
as  $\x_{\mu}^3 = (-1/y_6, y_2, -1/y_3)$ and the fourth one in the 
$\sigma_- \rightarrow \infty$ limit as $\x_{\mu}^4 = 
(y_6, -1/y_1, -1/y_4)$. For the $\ga = 0, \beta = \pi/4$ case the four
cusp positions reduce to $\x_{\mu}^1 = (-1,-1,1), \x_{\mu}^2 = (1,1,1),
\x_{\mu}^3 = (-1,1,-1), \x_{\mu}^4 = (1,-1,-1)$ which are cusp positions
of square Wilson loop (\ref{rf}). For the $\ga = \beta =0$ case the
four cusps degenerate to be two cusps specified by $\x_{\mu}^1 = 
(0,0,1), \x_{\mu}^3 = (0,0,-1)$ which are just two cusp positions
of (\ref{tc}), where the remaining two cusps $\x_{\mu}^2$ and
$\x_{\mu}^4$ are sended away to infinity.

Using the expressions in (\ref{yv}) we can show
\begin{equation} 
(\x_{\mu}^k - \x_{\mu}^{k+1})^2 = 0, \;\; k = 1, 2, 3, 4
\end{equation}
with $\x_{\mu}^5 = \x_{\mu}^1$, so that the four-cusp
Wilson loop indeed consists of four lightlike vectors.
In the projected $(\x_1, \x_2)$ plane
the square of the position vector $\tilde{\vx}^k$
 for the $k$-th cusp is larger than unity
\begin{equation}
(\tilde{\vx}^k)^2  = \frac{\vec{b}_k^2 \vec{b'}_{k}^2}
{(\vec{b}_k \cdot \vec{b'}_{k})^2} \ge 1, 
\label{xg}\end{equation}
where 
\begin{eqnarray}
\vec{b}_1 = (1, \tanh \ga),\; \vec{b'}_1 = (1, -\tan\beta),\; 
\vec{b}_2 = (1, \tanh \ga),\; \vec{b'}_2 = (1, \frac{1}{\tan\beta}),
\nonumber \\
\vec{b}_3 = (1, \tanh \ga),\; \vec{b'}_3 = (1, \tan\beta),\; 
\vec{b}_4 = (1, \tanh \ga),\; \vec{b'}_4 = (1, -\frac{1}{\tan\beta}).
\end{eqnarray}
The expression (\ref{xg})  shows the same factorized parametrization as
(\ref{fr}) for the hexagonal Wilson loop case.
Each cusp is located outside the unit circle and the figure of
the tetragonal Wilson
loop in the projected $(\x_1, \x_2)$ plane is characterized by
different values of $|\tilde{\vx}^k|, k =1,2,3,4$ and
$\tilde{\vx}^1\cdot \tilde{\vx}^2 = \tilde{\vx}^3\cdot \tilde{\vx}^4 = 0$.
The point $P_k$  between $\tilde{\vx}^k$ and 
$\tilde{\vx}^{k+1}$ defined by
$\mbox{\boldmath{$P$}}_k \cdot (\tilde{\vx}^k - \tilde{\vx}^{k+1}) = 0$
is specified by (\ref{tk}) with the parameter $t_k$ of (\ref{tx}).
Here we express $t_k$ as
\begin{equation}
t_1 = \frac{(\tan\beta + \tanh\ga)^2}{(1 + \tanh^2\ga)\sec^2\beta}, \;
t_2 = t_4 = \frac{1 - \tan^2\beta\tanh^2\ga}{(1 - 
\tanh^2\ga)\sec^2\beta}, \;
t_3 = \frac{(\tan\beta - \tanh\ga)^2}{(1 + \tanh^2\ga)\sec^2\beta}.
\label{tt}\end{equation}
The square of $\mbox{\boldmath{$P$}}_k$ given by (\ref{kp}) becomes
unity so that the four sides of the Wison loop are also tangent to
the unit circle. The time-component of the $k$-th point $P_k$ in the
$(\x_0, \x_1, \x_2)$ space is confirmed to be zero through (\ref{tt})
in the same manner as the hexagonal Wilson loop case.

\section{Conclusion}

Based on the asymptotic solution of the string with Euclidean worldsheet
in $AdS_3$ which was constructed by solving the axiliary linear problems
approximately associated with the generalized Sinh-Gordon model for the
field $\al$ in the complex $w$ plane \cite{FAM}, we have expressed the
approximate forms of the string solution
 near the cusps in the $AdS_4$ Poincare coordinates 
to extract the cusp positions in $R^{1,2}$ for the hexagonal and 
octagonal Wilson loops. A sequence of segments for these Wilson loops
living in $R^{1,2}$ have been confirmed to be described by the null
vectors. From these expressions of cusp positions we have observed that
the necessary conditions  for the hexagonal and octagonal 
Wilson loops to take closed contours are satisfied by using the relations
 for the Stokes matrices. 

We have demonstrated that the hexagonal Wilson loop is tangent to the
unit circle in the projected $(\x_1, \x_2)$ space and going alternating
up and down in $R^{1,2}$, where successive sign changings of cusp's
$\x_0$ are allowed only for the Wilson loop with an even number of
cusps. The asymptotic string configuration derived from the large
$w$ asymptotic solutions for the 
auxiliary linear left and right problems
in the $w$ plane has been confirmed to satisfy the string equations of
motion and the Virasoro constraints in the $z$ plane.

For the $\al = 0$ case we have solved the auxiliary  left and 
right linear problems in the $z$ plane 
to construct the general solutions
by making linear combinations of two independent solutions with
arbitrary coefficients. Owing to the reality and  
normalization conditions for the general solutions, the coefficients
for the left problem are characterized by appropriate comlex values
multiplying the trigonometrical functions of variable $\beta$,
while those for the right problem are expressed by the hyperbolic
functions of variable $\ga$. In the embedding coordinates $Y_{\mu}$
the former characterization turns out to be two SO(2) rotations
with angles  $\beta$ and $-\beta$ in the $(Y_{-1}, Y_0)$ plane
and the $(Y_1, Y_2)$ plane, while the latter one becomes two SO(2,4)
boosts with boost parameters $\ga$ and $-\ga$ in the $(Y_{-1}, Y_1)$
plane and the $(Y_0, Y_2)$ plane. 

We have demonstrated that the string
configuration with parameters $\beta$ and $\ga$ 
derived from the exact solutions of the linear problems
satisfies the equations of motion
and the Virasoro constraints for the string with Euclidean worldsheet
labelled by $(z,\bz)$ in the embedding coordinates $Y_{\mu}$.
Alternatively using the $AdS_3$ global coordinates $(t, \rho, \phi)$
we have confirmed that the $\beta, \ga$ string configuration
satisfies the equations of motion and the Virasoro constraints for the
string with Euclidean worldsheet labelled by $(\tau, \sigma)$.
The string solution with $\beta = \pi/4, \ga = 0$ is related to the
square Wilson loop with four cusps and the $\beta = \ga = 0$ string
solution is associated with the four semi infinite Wilson
lines with two cusps, where in this demonstration we note that 
two cusps are located at infinity.
In order to capture the figure of the Wison loop surrounding the
$\beta, \ga$ string surface we extract the asymptotic solution
near each cusp from the exact string solution. The four cusp 
locations in $R^{1,2}$  for the tetragonal Wison loop are determined
as functions of $\beta, \ga$ in the same way as the hexagonal and
octagonal Wilson loop cases by using only the asymptotic
solution. For the $\beta = \pi/4, \ga = 0$ string solution and
the $\beta = \ga =0$ one we have observed that the cusp positions
 determined from the asymptotic solution indeed 
agree with those fixed by the exact solution.


\begin{thebibliography}{99}
\bibitem{GKP} S.S. Gubser, I.R. Klebanov and A.M. Polyakov,
``A semi-classical limit of the gauge/string correspondence,"
Nucl. Phys. \textbf{B636} (2002) 99 [arXiv:hep-th/0204051].
\bibitem{AT} A.A. Tseytlin, ``Spinning strings and AdS/CFT duality,"
arXiv:hep-th/0311139; ``Semiclassical strings and AdS/CFT,"
arXiv:hep-th/0409296.
\bibitem{JP} J. Plefka, ``Spinning strings and integrable spin chains in
AdS/CFT correspondence," arXiv:hep-th/0507136.
\bibitem{AM} L.F. Alday and J. Maldacena, ``Gluon sacttering amplitudes
at strong coupling," JHEP \textbf{0706} (2007) 064 
[arXiv:0705.0303[hep-th]].
\bibitem{MK} M. Kruczenski, ``A note on twist two operators in 
$\mathcal{N}=4$ SYM and Wilson loops in Minkowski signature," 
JHEP \textbf{0212} (2002) 024 [arXiv:hep-th/0210115].
\bibitem{BDS} Z. Bern, L.J. Dixon and V.A. Smirnov, ``Iteration of planar
amplitudes in maximally supersymmetric Yang-Mills theory at three loops 
and beyond," Phys. Rev. \textbf{D72} (2005) 085001 [arXiv:hep-th/0505205];
C. Anastasiou, Z. Bern, L.J. Dixon and D.A. Kosower, 
``Planar amplitudes in maximally supersymmetric Yang-Mills theory,"
Phys. Rev. Lett. \textbf{91} (2003) 251602 [arXiv:hep-th/0309040].
\bibitem{AFK} S. Abel, S. Forste and V.V. Khoze, ``Scattering amplitudes
in strongly coupled $\mathcal{N}=4$ SYM from semiclassical strings in
AdS," JHEP \textbf{0802} (2008) 042 [arXiv:0705.2113[hep-th]];
E.I. Buchbinder, ``Infrared limit of gluon amplitudes at strong 
coupling," Phys. Lett. \textbf{B654} (2007) 46 [arXiv:0706.2015[hep-th]];
Z. Komargodski and S.S. Razamat, ``Planar quark scattering 
at strong coupling and universality," JHEP \textbf{0801} (2008) 044
[arXiv:0707.4367[hep-th]];
J. McGreevy and A. Sever, ``Quark scattering amplitudes at strong 
coupling," JHEP \textbf{0802} (2008) 015 [arXiv:0710.0393[hep-th]];
A. Popolitov, ``On coincidence of Alday-Maldacena-regularized
sigma-model and Nambu-Goto areas of minimal surfaces,"
arXiv:0710.2073[hep-th];
G. Yang, ``Comment on the Alday-Maldacena solution in calculating
scattering amplitude via AdS/CFT," JHEP \textbf{0803} (2008)
010 [arXiv:0711.2828[hep-th]];
Y. Oz, S. Theisen and S. Yankielowicz, ``Gluon scattering in deformed
$\mathcal{N}=4$ SYM," Phys. Lett. \textbf{B662} (2008) 297
[arXiv:0712.3491[hep-th]];
Z. Komargodski, ``On collinear factorization of Wilson loops and MHV 
amplitudes in $\mathcal{N}=4$ SYM," JHEP \textbf{0805} (2008) 019
[arXiv:0801.3274[hep-th]];
M. Kruczenski and A.A. Tseytlin, ``Spiky strings, light-like Wilson
loops and pp-wave anomaly," Phys. Rev. \textbf{D77} (2008) 126005
[arXiv:0802.2039[hep-th]];
R. Ishizeki, M. Kruczenski and A.A. Tseytlin, ``New open string 
solutions in $AdS_5$," Phys. Rev. \textbf{D77} (2008) 126018
[arXiv:0804.3438[hep-th]];
C.M. Sommerfield and C.B. Thorn, ``Classical worldsheets for string
scattering on flat and AdS spacetime," Phys. Rev. \textbf{D78} (2008)
046005 [arXiv:0805.0388[hep-th]].
\bibitem{ADI} D. Astefanesei, S. Dobashi, K. Ito and H. Nastase,
``Comments on gluon 6-point scattering amplitudes in $\mathcal{N}=4$
SYM at strong coupling," JHEP \textbf{0712} (2007) 077
[arXiv:0710.1684[hep-th]];
K. Ito, H. Nastase and K. Iwasaki, ``Gluon scattering in $\mathcal{N}=4$
super Yang-Mills at finite temperature," Prog. Theor. Phys. 
\textbf{120} (2008) 99 [arXiv:0711.3532[hep-th]];
S. Dobashi, K. Ito and K. Iwasaki, ``A numerical study of gluon 
scattering amplitudes in $\mathcal{N}=4$ super Yang-Mills theory at 
strong coupling," JHEP \textbf{0807} (2008) 088 
[arXiv:0805.3594[hep-th]];
S. Dobashi and K. Ito, ``Discretized minimal surface and the BDS 
conjecture in $\mathcal{N}=4$ super Yang-Mills theory at 
strong coupling," Nucl. Phys. \textbf{B819} (2009) 18
[arXiv:0901.3046[hep-th]].
\bibitem{LAM} L.F. Alday and J. Maldacena, ``Comments on gluon 
scattering amplitudes via AdS/CFT,"
JHEP \textbf{0711} (2007) 068 [arXiv:0710.1060[hep-th]];
A. Mironov, A. Morozov and T.N. Tomaras, ``On $n$-point
amplitudes in $\mathcal{N}=4$ SYM," JHEP \textbf{0711} (2007) 021
[arXiv:0708.1625[hep-th]]; ``Some properties of the Alday-Maldacena
minimum," Phys. Lett. \textbf{B659} (2008) 723 [arXiv:0711.
0192[hep-th]];
H. Itoyama, A. Mironov and A. Morozov, ``Boundary ring: a way to
construct approximate NG solutions with polygon boundary conditions:
I. $Z_n$-symmetric configurations," Nucl. Phys. \textbf{B808} (2009)
 365 [arXiv:0712.0159[hep-th]]; `` 'Anomaly' in $n = \infty$ 
Alday-Maldacena duality for wavy circle," JHEP \textbf{0807} (2008) 
024 [arXiv:0803.1547[hep-th]];
H. Itoyama and A. Morozov, ``Boundary ring or a way to
construct approximate NG solutions with polygon boundary conditions.
II. Polygons which admit an inscribed circle," Prog. Theor. Phys.
\textbf{120} (2008) 231 [arXiv:0712.2316[hep-th]];
A. Morozov, ``Alday-Maldacena duality and AdS Plateau problem,"
Int. J. Mod. Phys. \textbf{A23} (2008) 2118 arXiv:0803.2431[hep-th];
D. Galakhov, H. Itoyama, A. Mironov and A. Morozov, ``Deviation from
Alday-Maldacena duality for wavy circle," Nucl. Phys. \textbf{B823}
(2009) 289 [arXiv:0812.4702[hep-th]].
\bibitem{JKS} A. Jevicki, C. Kalousios, M. Spradlin and A. Volovich,
``Dressing the giant gluon," JHEP \textbf{0712} (2007) 047
 [arXiv:0708.0818[hep-th]];
A. Jevicki, K. Jin, C. Kalousios and A. Volovich, ``Generating AdS 
string solutions," JHEP \textbf{0803} (2008) 032
[arXiv:0712.1193[hep-th]].
\bibitem{KRT} M. Kruczenski, R. Roiban, A. Tirziu and A.A. Tseytlin, 
``Strong-coupling expansion of cusp anomaly and gluon amplitudes from
quantum open strings in $AdS_5\times S^5$," Nucl. Phys. \textbf{B791}
(2008) 93 [arXiv:0707.4254[hep-th]];
R. Roiban and A.A. Tseytlin, ``Strong-coupling expansion of
cusp anomaly from quantum superstring," JHEP \textbf{0711} (2007) 
016 [arXiv:0709.0681[hep-th]].
\bibitem{SR} S. Ryang, ``Conformal SO(2,4) transformations of the
one-cusp Wilson loop surface," Phys. Lett. \textbf{B659} (2008) 894
[arXiv:0710.1673[hep-th]];
``Conformal SO(2,4) transformations for the helical AdS string 
solution," JHEP \textbf{0805} (2008) 021 [arXiv:0803.3855]].
\bibitem{AR} L.F. Alday and R. Roiban, ``Scattering amplitudes, Wilson
loops and the string/gauge theory correspondence," Phys. Rept.
\textbf{468} (2008) 153 [arXiv:0807.1889[hep-th]].
\bibitem{CSV} F. Cachazo, M. Spradlin and V. Volovich, ``Iterative
structure within the five-particle two-loop amplitude," Phys. Rev. 
\textbf{D74} (2006) 045020 [arXiv:hep-th/0602228];
Z. Bern, M. Czakon, D.A. Kosower, R. Roiban and V.A. Smirnov,
``Two-loop iteration of five-point $\mathcal{N}=4$ super Yang-Mills
amplitudes," Phys. Rev. Lett. \textbf{97} (2006) 181601
 [arXiv:hep-th/0604074];  
Z. Bern, L.J. Dixon, D.A. Kosower R. Roiban, M. Spradlin,
C. Vergu and V. Volovich,`` The two-loop six-gluon MHV amplitude in
maximally supersymmetric Yang-Mills theory," Phys. Rev. 
\textbf{D78} (2008) 045007 [arXiv:0803.1465[hep-th]].
\bibitem{DHK} J.M. Drummond, J. Henn, G.P. Korchemsky and E. Sokatchev,
``On planar gluon amplitudes/Wilson loops duality," Nucl. Phys. 
\textbf{B795} (2008) 52 [arXiv:0709.2368[hep-th]];
``Conformal Ward identities for Wilson loops and a test of the duality
with gluon amplitudes," arXiv:0712.1223[hep-th];
``The hexagon Wilson loop and the BDS ansatz for the six-gluon 
amplitudes," Phys. Lett. \textbf{B662} (2008) 456 
[arXiv:0712.4138[hep-th]];
``Hexagon Wilson loop $=$ six-gluon MHV amplitude," 
Nucl. Phys. \textbf{B815} (2009) 142 [arXiv:0803.1466[hep-th]];
C. Anastasiou, A. Brandhuber, P. Heslop, V.V. Khoze, B. spence and 
G. Travaglini, ``Two-loop polygon Wilson loops in $\mathcal{N}=4$ 
SYM," arXiv:0902.2245[hep-th].
\bibitem{KP} K. Pohlmeyer, ``Integral Hamiltonian systems and 
interactions through quadratic constraints," Commun. Math. Phys.
 \textbf{46} (1976) 207.
\bibitem{VS} H.J. de Vega and N. Sanchez, ``Exact integrability of 
strings in D-dimensional de Sitter spacetime," Phys. Rev. \textbf{D47}
  (1993) 3394;
M. Grigoriev and A.A. Tseytlin, ``Pohlmeyer reduction of $AdS_5\times
S^5$ superstring sigma model," Nucl. Phys. \textbf{B800}  (2008) 450
[arXiv:0711.0155[hep-th]];
``On reduced models for superstrings on $AdS_n\times S^n$,"
Int. J. Mod. Phys. \textbf{A23} (2008) 2107 [arXiv:0806.2623[hep-th]];
J.L. Miramontes, ``Pohlmeyer reduction revisited," JHEP \textbf{0810} 
(2008) 087 [arXiv:0808.3365[hep-th]].
\bibitem{JJ} A. Jevicki and K. Jin, ``Solitons and AdS string 
solutions," Int. J. Mod. Phys. \textbf{A23} (2008) 2289 
[arXiv:0804.0412[hep-th]];
``Moduli dynamics of $AdS_3$ strings,"  JHEP \textbf{0906} (2009) 064
[arXiv:0903.3389[hep-th]].
\bibitem{DJW} H. Dorn, G. Jorjadze and S. Wuttke, ``On spacelike and
timelike minimal surfaces in $AdS_n$," arXiv:0903.0977[hep-th];
H. Dorn, ``Some comments on spacelike surfaces with null polygonal
boundaries in $AdS_m$," arXiv:0910.0934[hep-th].
\bibitem{GJ} G. Jorjadze, ``Singular Liouville fields and spiky strings
in $R^{1,2}$ and SL(2,R)," arXiv:0909.0350[hep-th].
\bibitem{FAM} J.F. Alday and J. Maldacena, ``Null polygonal Wilson 
loops and minimal surfaces in Anti-de-Sitter space,"
arXiv:0904.0663[hep-th].
\bibitem{GMN} D. Gaiotto, G.W. Moore and A. Neitzke, ``Four-dimensional
wall-crossing via three-dimensional field theory," 
arXiv:0807.4723[hep-th];
``Wall-crossing, Hitchin systems, and the WKB approximation,"
arXiv:0907.3987[hep-th].
\bibitem{SS} K. Sakai and Y. Satoh, ``A note on string solutions in
$AdS_3$," HEP \textbf{0910} (2009) 001 [arXiv:0907.5259[hep-th]].



\end{thebibliography}
\end{document}